\documentclass[11pt,tightenlines]{revtex4-2}

\usepackage[utf8]{inputenc}
\usepackage[OT4]{fontenc}

\usepackage{amsmath,upgreek}
\usepackage{amssymb}
\usepackage{epsfig}

\begin{document}


%


%

\title{The carnivorous plant {\it Genlisea} harnesses active particle dynamics to prey on microfauna}

\author{Jos{\' e} Mart{\' i}n-Roca, C. Miguel Barriuso-Guti{\' e}rrez, Ra{\' u}l Mart{\' i}nez Fern{\' a}ndez, Camila Betterelli Giuliano,\\ Rongjing Zhang, Chantal Valeriani \& Laurence G. Wilson}
\date{\today}

\maketitle
\section*{Abstract}
Carnivory in plants is an unusual trait that has arisen multiple times, independently, throughout evolutionary history.  Plants in the genus {\it Genlisea} are carnivorous, and feed on microorganisms that live in soil using modified subterranean leaf structures (rhizophylls).  A surprisingly broad array of microfauna has been observed in the plants’ digestive chambers, including ciliates, amoebae and soil mites.  Here we show, through experiments and simulations, that {\it Genlisea} exploit 
active matter physics to ‘rectify’ bacterial swimming and establish a local flux of bacteria through the structured environment of the rhizophyll towards the plant’s digestion vesicle.  In contrast, macromolecular digestion products are free to diffuse away from the digestion vesicle and establish a concentration gradient of carbon sources to draw larger microorganisms further inside the plant.  Our experiments and simulations show that this mechanism is likely to be a localised one, and that no large-scale efflux of digested matter is present.
\clearpage

Carnivorous plants are unusual organisms that have evolved to survive in nutrient-poor environments by trapping and digesting animal prey, typically insects.  Most commonly this is to supplement their intake of soil macronutrients such as nitrogen and phosphorus \cite{Ellison_PlantBiol06,AlbertRJTMDT_JXB10}.  Carnivorous plants have adopted a range of prey-capture strategies: pitfall traps (\textit{Sarracenia, Nepenthes, Heliamphora}), `flypaper' traps (\textit{Drosera, Pinguicula}) and suction traps (\textit{Utricularia}) amongst others.  Carnivory has also given insight into evolutionary biology.  Not only is carnivory a rare trait, but there are several evolutionarily distinct lineages that have arrived at the same basic trapping principle --- for example, the sticky-leaved flypaper traps of \textit{Byblis} and \textit{Drosophyllum} \cite{AlbertSWMC_Science92}.  The genus \textit{Genlisea} is comparatively obscure, though it was included in Darwin's work `Insectivorous Plants' in 1875 \cite{Darwin_insect}.  Approximately 30 extant species are distributed across tropical Africa, Central and South America, often in inaccessible and sparsely populated regions.

\begin{figure}[pht]
\begin{center}
\includegraphics[width=1\linewidth]{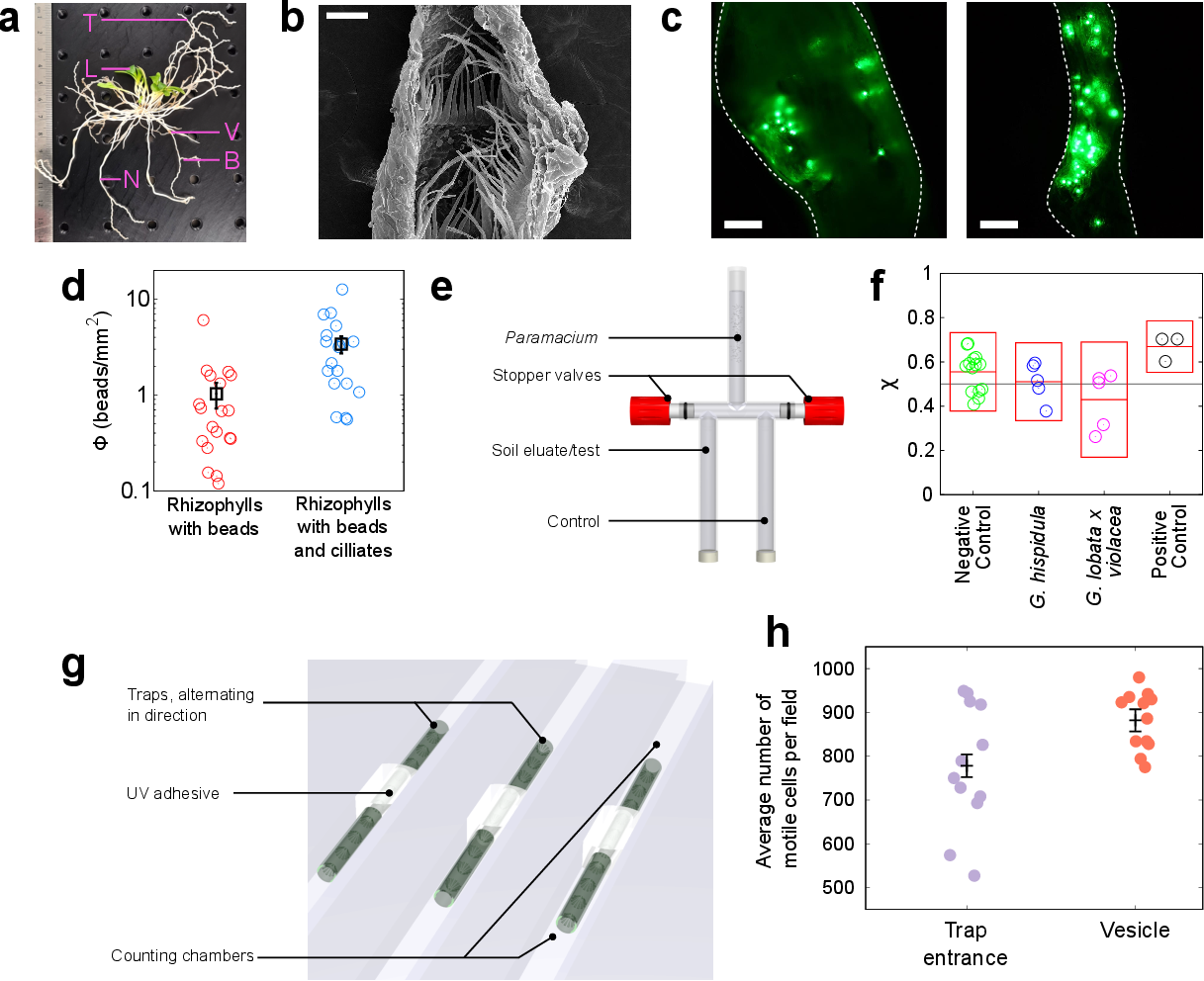}
\end{center}
\caption{{\bf Experiments showing transport of active and inactive matter.} {\bf a} Photograph of a \emph{G. hispidula} plant:  Photosynthetic leaf \emph{(L)}, digestive vesicle \emph{(V)}, trap neck \emph{(N)}, trap bifurcation \emph{(B)}, and the characteristic `corkscrew' structure containing the trap openings \emph{(T)}.  {\bf b}  A cut-away SEM image showing the interior of one of the traps.  The detentive hairs point upwards along the rhizophyll's central channel towards the digestive vesicle (100\,$\mu$m scale bar).  {\bf c}  An epi-fluorescence image of 15 $\mu$m  tracers  inside the  vesicle (left) and the trap neck (right) of a \emph{G. hispidula} rhizophyll. {\bf d} Concentration of  15\,$\mu$m fluorescent tracers (normalized by the trap's open area) in the absence (red) or presence (blue) of ciliate preys (mean values in black, errors bars show S.E.M.)  {\bf e} Chemotaxis assay chamber presenting different stimuli to planktonic ciliates (see text).  {\bf f}  Chemokinesis coefficients.  Each point corresponds to an experiment involving an initial loading of around 150 ciliates.  Red boxes indicate 95\% confidence intervals.  The horizontal black line represents the null hypothesis. {\bf g} A `cartoon' of the bacterial rectification setup, with a section of rhizophyll trap neck connecting two chambers.  {\bf h}  Number of cells at each end of the trap neck section after two hours (black data points represent mean and 95\% CI).}
\label{Fig1}
\end{figure}

The parts of \textit{Genlisea} spp. that lie above the surface are superficially unremarkable.  The plant is marked by a rosette of small oblong or obovate green leaves, up to a centimeter or so in size.  Extending beneath the soil from the main core of the plant are a series of white or translucent tube-like rhizophylls that possess a pronounced bulge (the vesicle) part-way along their length, and split into two twisted terminal structures (see Fig. \ref{Fig1}a).  From an anatomical point of view, {\it Genlisea} is rootless \cite{Fleischmann_book}; these white appendages are underground leaves specially adapted for the capture and digestion of microorganisms.  The interior of the rhizophyll is hollow from the vesicle to the spiral-shaped openings at its distal end.  The hollow core is filled with rows of detentive hairs that point upwards towards the vesicle; these have been posited to act like an eel or lobster trap \cite{PlachnoMK-KPS_AnnBot07}, allowing soil-dwelling organisms such as soil mites to pass inwards while making escape difficult.  Glands, most densely clustered in the rhizophyll vesicle, secrete digestive enzymes \cite{Heslop-Harrison_book,PlachnoLAILMPWAJV_PlantBiol06} which break down prey, and reabsorb nutrients through pores in the cuticle of the digestive glands.  The interior milieu of the vesicle has been characterised as mucilaginous \cite{PlachnoJFAJ_ActaBotGall05}.

The manner in which \textit{Genlisea} traps its prey remains controversial.  Darwin noted that microorganisms entering the branches of the rhizophylls would find their egress prevented by the rows of detentive hairs, but stated that it is not clear what would entice the microorganisms to enter in the first place.  Furthermore, he noticed that the digestive vesicles of the plant were filled with soil particles and other inorganic debris; this debris is often seen in older rhizophylls, and is too large to have arrived there by Brownian motion.  Juniper \emph{et al.} \cite{Juniper_book} suggest that the plants actively pump fluid into their traps, similar to the closely related bladderworts (\emph{Utricularia} spp.), although at that time there were few studies in living plants, and the flow rates and concomitant energy consumption required to sustain flow seem prohibitive.   Several authors have speculated upon whether the plant uses an attractant to lure prey into its traps.  Barthlott \emph{et al.} introduced ciliates tagged with $^{35}$S and later found the tags had accumulated in the leaves of the plant \cite{BarthlottSPEFBG_Nature98}.  Darnowski and Fritz \cite{DarnowskiSF_CPN10} tested whether agar that had been placed near {\it Genlisea} had absorbed any putative `lure' chemical.  The production of lure molecules will incur an energetic cost on the plant, discouraging their creation.  In this work, we show that active matter physics principles play a hitherto unrecognised role in the flux of living matter into {\it Genlisea}.

Over recent decades, Soft Matter Physics and Statistical Physics have given useful insights to unravel features of complex biological phenomena such as cell division, tissue morphogenesis and population dynamics \cite{Petridou2021,Mashaghi2014,Allen2019}.  Active Matter represents a fundamentally new non-equilibrium branch of soft condensed matter physics, studying out-of-equilibrium systems in which energy is supplied at the level of individual entities and translated into unidirectional motion, for example `active particles' \cite{Ramaswamy2010}. These  dissipate energy while moving
\cite{Ramaswamy2017,Bechinger2016}. Active systems give rise to unexpected collective phenomena not observed in equilibrium systems,  from colonies of bacteria to flocks of birds \cite{Vicsek2012,Marchetti2013,Bechinger2016}.  Active particles can be synthetic (such as active colloids \cite{Yan2016,vanKesteren2023,Calero2020,vanderLinden2020,Vutukuri2020,Palacci2014}) or living (such as bacteria \cite{Manson2020,Cates2013,Qiu2014}).
Microorganisms live in  environments where viscous forces are orders of magnitude larger than inertial forces (\textit{i.e.} low Reynolds numbers environments).  Here, fluid motion is described by the time-independent Stokes equation.  This gives rise to behaviour qualitatively different to the macroscopic case, and so analogies drawn between \textit{Genlisea} and human-scale eel traps or lobster pots are only superficial --- different physics obtains at the microscale.  Bacterial cells (sizes $\sim 10^{-6}$\,m) often swim in a series of relatively straight runs (length $\sim 10^{-4}-10^{-5}$\,m) separated by reorientation events.  Depending on the species, bacterial reorientations vary from deflections in swimming trajectory \cite{BergDB_Nature72} through to reversals \cite{ThorntonJBSDBBLW_NatComms20,XieTASCX-LW_PNAS10,SonJGRS_NatPhys13} or complete stops \cite{ArmitageRS_Microbiol97}, during which time Brownian motion reorients cells.  One of the most commonly observed swimming phenotypes is the `run-and-tumble' motility observed in soil bacteria such as {\it Bacillus subtilis} and enteric bacteria such as {\it Escherichia coli}.  When {\it E. coli} is confined in the presence of a microfabricated wall, Galajda and coworkers demonstrated that its motion is rectified by  funnel-shaped openings  \cite{galajda2007wall}.  This behaviour is qualitatively different to that of true Brownian particles \cite{ReichardtCR_PRL06} and from that of active polar particles\cite{Martinez2018} in the same confining geometry.

We find that {\it Genlisea} exploits phenomena observed in the study of active matter in the presence of obstacles  to capture prey.  We demonstrate  how an unusual carnivorous plant has harnessed the  rectification of bacterial motion  to capture microorganisms, while allowing an attractive chemical gradient  of organic molecule to form inside its traps.  We first demonstrate that the presence of larger prey microorganisms enhanced transport of environmental debris into the traps, but show no evidence for a chemical `lure' for these organisms.  Next we demonstrate that {\it Genlisea} can rectify bacterial swimming,  that the geometry of the hairs within the plants is close to the optimal value for trapping, and that the hairs (rather than a proteinaceous mucilage) are likely the dominant contributor to trapping.

Figure \ref{Fig1}.a show the gross morphology of an excavated plant, and Figure \ref{Fig1}.b  a scanning electron microscopy cross-section of a trap, showing the detentive hairs.  To establish a mechanism for large debris to arrive in the vesicle under normal growth conditions ({\it rhizophylls} grow downward, so material is transported against gravity), we used fluorescent particles, diameter 15$\upmu$m, as probes (Fig. \ref{Fig1}.c).   Excised but intact {\it rhizophylls} were placed in a sample chamber containing water and probe particles, both with (right-hand side panel) and without (left-hand side panel) ciliate microorganisms ({\it Paramecium multimicronucleaton}, with cell bodies around 100 $\upmu$m long and 50 $\upmu$m in diameter).  
To compare rhizophylls of different sizes,  we normalised the data by dividing the number of beads counted in a trap ($N_{beads}$) by the area of the trap `mouth' ($A_{mouth}$) to give a net flux of beads, as shown in  Fig. \ref{Fig1}.d: $\Phi=N_{beads}/A_{mouth}$.  
The chosen particles were too large to be significantly transported by Brownian motion in the vertical direction.
The presence of ciliates ({\it Paramecium multimicronucleaton}) gives a roughly three-fold increase in the number of particles in the vesicles (Fig. \ref{Fig1}.c and d).  Therefore,  the presence of microorganisms is  sufficient to transport inorganic material into the traps.  We note that placing rhizophylls in the sample chambers, and removing them for particle counting causes a small flow within the sample chamber, which appears to be responsible for some transport, albeit it at a lower level: a few particles could be found in the rhizophylls irrespective of the presence of microorganisms.  

To investigate the presence or absence of a prey attractant, we built a T-maze choice assay of the type used by van Houten \emph{et al.} \cite{vanHoutenEMTK_JProtozool82} (diagram in Fig. \ref{Fig1}.e), and tested the reaction to soil eluate of ciliates similar to those previously determined to be prey microorganisms \cite{BarthlottSPEFBG_Nature98}.  Eluate from a pot containing a plant was placed in one test arm of the T-maze, and a control solution in the other.  Negative controls were provided by eluate from bare media pots, and by placing eluate in both arms of the maze.  A 5 mM solution of NH$_4$Cl was used as a positive control \cite{vanHoutenEMTK_JProtozool82}.  5 ml of distilled water containing \emph{P. multimicronucleaton} were placed in the final arm of the T-maze, and the stopcock opened.  After 30 mins, the stopcock was closed and the ciliates in each arm were counted under a microscope at low magnification and dark field illumination.  The effect of the chemical stimulus was measured using the `index of chemokinesis' ($\chi$), reported in  Fig. \ref{Fig1}.f: 
\begin{equation}
\chi = \frac{N_{test}}{N_{test}+N_{control}},
\label{Chemokin}
\end{equation}
where $N_{test}$ and $N_{control}$ are the numbers of cells in the test and control arm of the T-maze, respectively.  A value of $\chi=1$ indicates a chemoattractant strong enough to draw all cells into the test arm of the choice chamber, and a value of $\chi=0$ corresponds to a perfect repellent (all cells in the control arm).  A value of $\chi=0.5$ corresponds to the null result: the microorganisms are neither attracted or repelled by the test substance (shown by a continous line in  Fig. \ref{Fig1}.f).   The values of $\chi$ reported in Fig. \ref{Fig1}.f show no clear evidence of a chemical lure for these ciliates.
  
Previous studies of the carnivory of {\it Genlisea} have focused on interactions with protozoa, but many smaller microorganisms exist in the soil alongside them.  In a nutrient-poor environment, a broader prey spectrum increases the chances that nutritional requirements will be met.  It has been estimated that there are between $10^4$--$10^5$ protozoa per gram of soil \cite{FOISSNER2014}, compared to around $10^8$ bacteria in the same mass \cite{RaynaudNN_PLoSOne14}.  The biomasses of these categories are therefore likely to be similar.  Rhizosphere microbiomes are complex \cite{LundbergSLSPSYJGSMJTAEVKTdRRETERLPHSTJD_Nature12}, but the swimming phenotype of common model soil-dwelling and root-associated bacterial species such as {\it B. subtilis} is a canonical run-and-tumble quantitatively similar to {\it E. coli} \cite{TurnerLPMNHB_BiophysJ16}.  The size and arrangement of detentive hairs within the rhizophylls is reminiscent of microfabricated devices used in previous studies of active matter systems \cite{galajda2007wall,ReichhardtCR_PRE04,Martinez2020}.  Inspired by such results, we prepared chambers divided into two, with the halves bridged by excised rhizophylls (shown in `cartoon' form in Fig. \ref{Fig1}g).  The rhizophylls were aligned in opposite directions in alternating channels, and both halves of the chambers filled with an initially uniform concentration of {\it E. coli}.   After two hours, bacteria in a 2 mm$^2$ field of view at either end of each rhizophyll were counted. As shown in Figure 1.h we observed an enrichment of 10--15\% in the number of cells present at the end of the rhizophyll previously connected to the vesicle (as compared to the trap entrance).

\begin{figure}[pht]
\begin{center}
\includegraphics[width=1\linewidth]{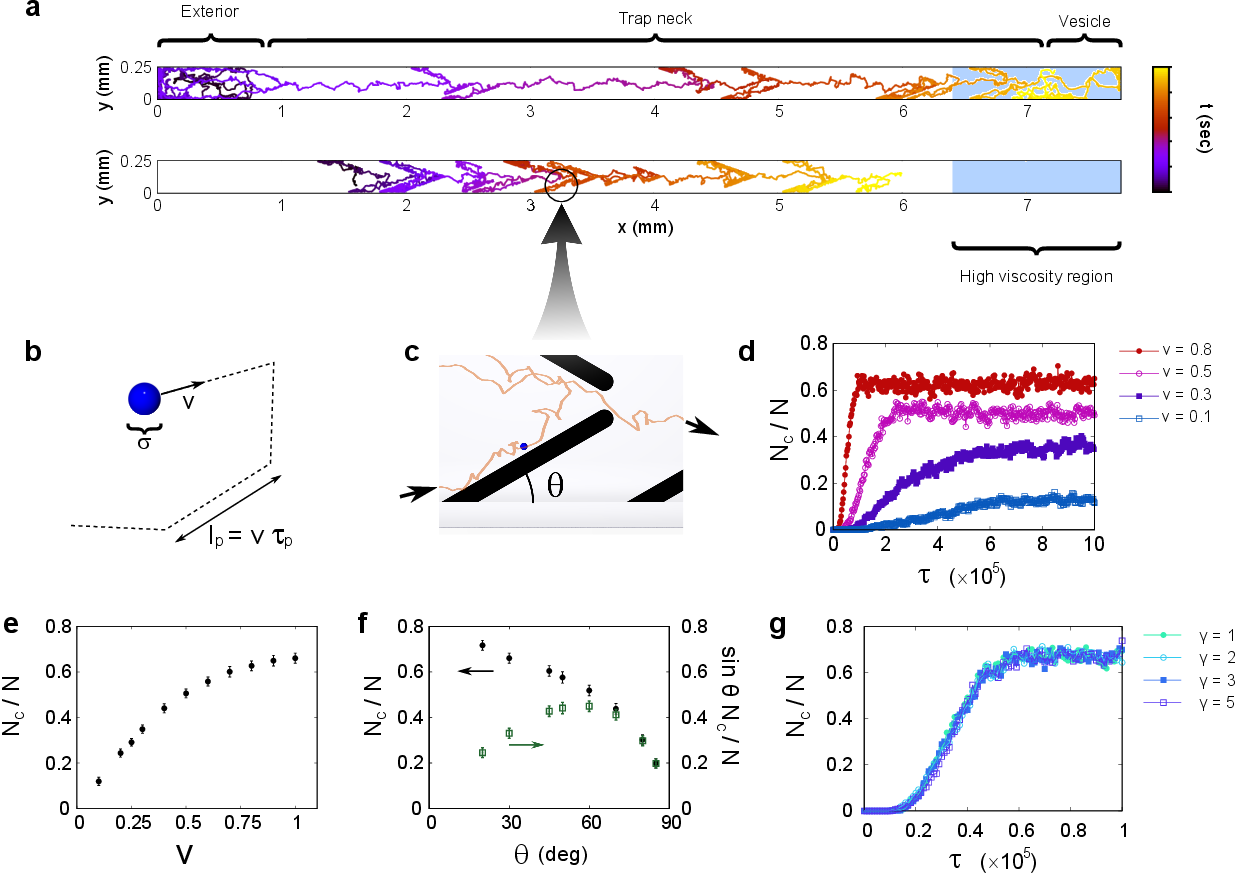}
\end{center}
\caption{ {\bf Geometrical contributions to carnivorous trapping.} {\bf a} Trajectories of simulated bacteria within the rhizophyll of {\it Genlisea}, with dimensions taken from experiments.  The detentive trap hairs have been omitted for clarity.  The structures have three regions including an `exterior' (no hairs), neck (with hairs), and vesicle (no hairs, but variable friction).  {\bf b} A `cartoon' showing characteristic swimming mode and parameters for the bacterium: size ($\sigma$), speed ($v$) persistence time ($\tau_p$) and persistence length ($l_p=v\tau_p$). {\bf c} Subsection of a particle trajectory from panel {\bf a} showing a simulated bacterium interacting with a hair.  The hair lies at an angle $\theta$ to the rhizophyll wall; black arrows show the direction of the bacterium's travel.  {\bf d}  The trapping efficiency $N_c/N$ for fixed trap geometry 
and variable swimming speed. Both trapping rate (initial gradient) and saturation occupancy (plateau level at late times) increase with $v$; the time is indicated in non-dimensionalised simulation units.  {\bf e} The saturation level of $N_c/N$ as a function of swimming speed.  The escape probability decreased with $l_p$. {\bf f}  Dependence of trapping efficiency on the hair angle $\theta$, for hairs with a fixed projection on the $y$-axis (see text).  Smaller angles lead to more efficient trapping, but at a higher cost of production.  A first-order correction to these values that takes into account the extra cost of producing longer hairs is shown with green squares (right-hand vertical axis), and peaks around 60$^\circ$.  {\bf g} The effect of decreased mobility within the vesicle on trapping efficiency.  There is no apparent evidence that increasing the friction coefficient in the vesicle increases trapping; the hairs are therefore the dominant contribution to prey trapping.}
\label{Fig2}
\end{figure}

We use this information to guide numerical simulations that untangle the relative importance of different aspects of the trap structure.  We simulated suspensions of  run-and-tumble particles (disk-like, with diameter $\sigma$ and propulsion speed of $v$) confined in a channel.  Figure \ref{Fig2}a represents typical trajectories of active particles in the funneled channel,  mimicking the plant's hair and whose geometry has been tailored borrowing parameters from experiments: the individual tracks are colour-coded to indicate time.  Detentive hairs within the traps have been omitted for clarity, but their influence can be seen in the chevron-shaped deviations in the trajectories, which lead to the trap vesicle on the right. The particles' motion is characterised by a persistence length $l_p=v\tau_p$, where $\tau_p$ is the persistence time, or the time between reorientations (Fig. \ref{Fig2}.b).  Interactions with the detentive hairs are shown in Fig.\ref{Fig2}.c, where $\theta$ is the angle that the hairs make with the rhizophyll wall.  Two-dimensional numerical simulations were carried out (see Supplementary Materials).

 To quantify the trapping efficacy, we examine the rate of accumulation in the vesicle, as well as the final (steady state) fraction of cells located there.  Figure \ref{Fig2}.d shows the time dependence of cell accumulation, for cells with different swimming speeds ($v=0.1$--$0.8$, as indicated in the legend).  Intuitively, cells that swim faster accumulate more quickly in the vesicle.  The `toy model' invoked is clearly sufficient to allow trapping of bacteria in the vesicle:  the vesicle comprises 10\% of the length of the rhizophyll, but accommodates at least 15\% of the cells, even in the case of the weakest trapping, rising to over 60\% of cells for the strongest trapping.   If the cells' tumble rate is held constant but the speed is varied, this will give rise to a longer persistence length $l_p$.  This results in a more efficient rectification of the swimming behaviour, and therefore a greater fraction of cells accumulating within the vesicle.

 Nevertheless, the stochastic swimming process does offer cells a chance of escape.  Figure \ref{Fig2}.e shows the steady-state accumulation of cells in the vesicle, as a function of swimming speed.  The accumulation saturates with around 65\% of the cells in the vesicle, presumably limited by cells stochastically escaping the trap.  Our microscopy studies, as well as those of other authors \cite{Fleischmann_book}, show that the detentive hairs within the rhizophyll vary in length somewhat between species, but that the gap in the rhizophyll centre is somewhat constant.  We therefore chose to vary $\theta$ and the hair length $h$, while keeping $h \text{sin} \theta$ (their projection on the $y$-axis) constant.  Figure \ref{Fig2}.f shows that the trapping efficiency, expressed as the ratio of the number of trapped cells ($N_c$) to the total number of cells ($N$) increases as $\theta$ decreases (filled symbols), up to a saturation level that is close to the value achieved by increasing the swimming speed (or persistence length).  Although smaller values of $\theta$ will lead to increased trapping, there is an energetic cost associated with growing longer hairs, which to a first approximation scales linearly with $h$.  For a constant value of $h \text{sin} \theta$, the trapping efficiency becomes $N_c \text{sin} \theta /N$, which is plotted against the second vertical axis of \ref{Fig2}.f (empty symbols), showing a peak at around 60$^\circ$, close to the value observed in SEM studies.  
    
 Lastly, we investigated the effect of the mucilaginous plug that has been observed in the vesicle of species of {\it Genlisea}.  It is difficult to determine definitively whether the mucilage is a product of the plant, or the consequence of the digestion of microbes.  Nevertheless, its rheology may have consequences for the trapping behaviour of the plant.  In a situation where the viscosity is variable, we anticipate that purely Brownian diffusing particles will have a distribution $\rho(r)$ that is uniform, because $\rho(r)\sim \text{exp}\left[{-\frac{U(r)}{k_BT}} \right]$ at equilibrium.  Conversely, swimming particles accumulate in regions of lower mobility.   We therefore perform simulations in which the stomach region has increased friction $\gamma$, Fig. \ref{Fig2} g.  These show that the effect of lower particle mobility has only a marginal trapping effect, and that the accumulation in the vesicle is governed by the geometry of hairs in the rhizophyll channel.

We have shown that plants in this fascinating genus constitute a naturally-occurring active matter rectifier, allowing them to increase their supply of nutrients in an otherwise nutrient-poor environment by trapping bacterial prey.  Darwin's `eel trap' description \cite{Darwin_insect} was based on observations of microfauna that became stuck in {\it Genlisea} rhizophylls, unable to escape.  We show that the same structures that trapped these arthropods are sufficient to guide bacteria --  orders of magnitude smaller -- to the plant's digestive vesicles.  Moreover, we explain the presence of large soil particles in the vesicle without recourse to an active mechanism such as fluid flows (for which there is little evidence in this genus \cite{PlachnoKAJFJK_ActaBotGall05}), but find no evidence for a chemical lure, at least one suitable for attracting a common genus of ciliate.  Although this plant is considered a `true' carnivore due to its production of protease \cite{Heslop-Harrison_Chapter}, carnivory is a spectrum, with genera such as {\it Roridula} using symbiosis with insect commensals to facilitate prey digestion \cite{GlisteningCarnivores}.  Our study raises the intriguing possibility of non- or quasi-carnivorous subterranean structures in a potentially wide range of other plants that use quirks of their morphology to sequester microorganisms that then die near to the plant, releasing nutrients --- a web of subtle carnivory beneath our feet.


\section*{Materials and Methods}
\subsection*{Plant Growth}

Experiments were conducted on \emph{Genlisea hispidula}, with additional measurements on \emph{Genlisea lobata x violacea} in the case of the chemoattractant tests.  The plants were grown in 5 cm pots filled with a 1:1:1 peat-sand-perlite soil medium.  Each pot was placed into an individual plastic cup within the terrarium to prevent the water outflow from each pot mixing together (necessary for the chemotaxis experiments).  The plants were watered with distilled water, and placed on a 16 hour photoperiod under `cool white' compact fluorescent lamps (color temperature 6500 K).  During cultivation, a USB data logger was used to record the average temperature and relative humidity in the terrarium: around 24$^\circ$C and 45\% humidity during the day, and 19.5$^\circ$C at night with 65\% humidity.  The microscopy experiments were conducted at the ambient temperature of the lab (away from the fluorescent lighting) of 21$\pm1^\circ$C.  The SEM image in Fig. \ref{Fig1}b is from {\it G. hispidula}, prepared according to the protocols provided in previous studies \cite{Juniper_book,PlachnoJFAJ_ActaBotGall05,Fleischmann_book}.

\subsection*{Transport of debris into vesicles}
As previously stated, inorganic debris is found inside the digestive vesicle of \emph{Genlisea} spp., and this debris must be actively moved into the traps against gravity.  The most likely agent seems to be the prey animals that live in the surrounding environment.  To test the hypothesis that prey animals push inorganic material into the traps, several plantlets of \emph{G. hispidula} were excavated and washed thoroughly with distilled water.  Around six rhizophylls were excised from the plants and cut above the digestive vesicle so that the hollow rhizophyll channel was maintained.  The plants and excised traps were placed in separate petri dishes, which were divided into two groups, test and control.  Petri dishes in the test group contained a suspension of fluorescent beads (diameter 15 $\upmu$m, `dragon green', Bangs Labs Inc.) mixed with a culture of the ciliate \emph{Paramecium multimicronucleaton} (Carolina Biological Supply Co.); the control group of petri dishes contained only a suspension of beads (at the same concentration as the test dishes).  The particles that we use are too large to be significantly transported by Brownian motion in the vertical direction; we would expect their concentration do decrease exponentially against gravity, with the characteristic length scale given by Perrin \cite{Perrin_book}, $l_c=k_BT/m^*g$, where $k_BT$ is the thermal energy, $m^*$ is a particle's buoyant mass and $g$ is acceleration due to gravity.  For particles the size of ours, $l_c \approx 5 \mu$m, so we expect the fluorescent tracers to lie more or less in a layer on the bottom surface of the petri dish, though they are free to diffuse in the horizontal plane.

After 7 days, plants and traps were washed extensively with distilled water and analyzed under a fluorescence microscope.  The number of beads found inside each trap was recorded.  As can be seen in Fig. \ref{Fig1}a, the rhizophylls from a single plant are often of different sizes, with different trap thicknesses and lengths.  To make a valid comparison between rhizophylls, we normalize the number of beads each trap ingested by its effective opening (`mouth') size, giving a total flux of $\Phi$ beads/mm$^2$.  The trap shapes are rather complicated, so we modeled each trap arm as a cylinder with a opening fraction per unit surface area, and the `mouth' region at the trap bifurcation as an open rectangle.  The dimensions of the trap were determined by digital photographs. 

\subsection*{Presence of a chemoattractant}
To test for the presence or absence of a soluble chemoattractant produced by the plant, we initially attempted some holographic particle tracking \cite{WilsonRZ_OptExp12,ThorntonJBSDBBLW_NatComms20} to see if the swimming patterns of \emph{P. multimicronucleaton} were modified when swimming close to the rhizophyll.  These results were inconclusive so instead we performed a T-maze choice assay of the type used by van Houten \emph{et al.} \cite{vanHoutenEMTK_JProtozool82} (as in Fig. \ref{Fig1} {\bf e}).  Five pots containing \emph{G. hispidula} plants, five containing \emph{G. lobata $\times$ violacea}, and five pots containing bare potting media without {\it Genlisea} were kept in the conditions described above.  Each pot was placed in an individual plastic cup so that the water outflow from the pots could not mix together.  After 3 months, each pot was washed through with around 50 ml of distilled water, and the eluate that accumulated in the plastic cups was collected.  The eluate from a pot containing a plant was placed in one test arm of the T-maze, and a control solution in the other.  Negative controls were provided by the eluate from bare media pots, and by comparing plant solutions to themselves.  A 5 mM solution of NH$_4$Cl was used as a positive control \cite{vanHoutenEMTK_JProtozool82}.  5 ml of distilled water containing \emph{P. multimicronucleaton} was placed in the final arm of the T-maze, and the stopcock opened.  After 30 mins, the stopcock was closed and the ciliates in each arm were counted under a microscope at low magnification and dark field illumination.  The effect of the chemical stimulus was measured using the index of chemokinesis $\chi$ (Eq. \ref{Chemokin}).  

\subsection*{Bacterial rectification}
To demonstrate that the rhizophyll `neck' is capable of rectifying bacterial swimmers, an assay was developed as pictured in Fig. \ref{Fig1}g.  Channels measuring 50 mm $\times$ 4mm $\times$ 3 mm were constructed from UV-curing glue and glass slides.  2 cm sections of trap `necks' cut from {\it G. hispidula} were carefully washed in DI water and placed in the channels.  These were sealed in place using UV curing glue to give an external barrier between the ends of the neck sections.  The chambers were then filled with a suspension of {\it E. coli} bacteria in tryptone broth as described previously \cite{WilsonVMJS-LJTGBPPWP_PRL11}.  The suspension initially occupied both halves of the chamber, with bacteria at a uniform concentration.  After a period of 2 hours, the density of bacteria immediately adjacent to the neck openings was measured.

\subsection*{Simulation details}


To explore the hypothesis of living particles being trapped inside the rhizophylls, we numerically study the relationship between the motion of  prey organisms and the  hair geometry. Based on the experimental observation of \cite{galajda2007wall}, our model consists of a two dimensional system of  $N$ run-and-tumble particles. The active particles of diameter $\sigma$ are kept in a two-dimensional closed geometry which consists of a channel with hairs at regular distances, of identical shape and orientation. Channel hairs are composed of non-motile circles, while the channel's walls  are directly simulated as straight boundaries.This geometry (shown in `cartoon' form in Fig. \ref{Fig2}a), is a simplified representation of a rhizophyll. 
Particles move due to self-propulsion in a straight line at speed $v$ during a period of time $\tau_p$, after which they randomly change their direction of motion, but not their speed. Interactions between particles and with the walls of the channel are solved via a Molecular Dynamics algorithm computing the interacting forces between particles using a WCA potential (see Supplementary Material for more details). We consider a channel of width $Ly=25\sigma_0$ and length $L_x=775\sigma_0$, containing particles of diameter $\sigma_0$ at a number density $\rho=0.1 \sigma_0^{-2}$. The distance between consecutive hairs along the walls is $d=25\sigma_0$. The opening between opposing hairs in the channel is $3\sigma_0$. In Fig 2.e the values used for the simulations are $\theta=60^\circ$ (tilt angle of the hairs respect to the longitudinal axis of the channel), $v=1.0 \sigma_0/\tau_0$, $\tau_r= 1.0 \tau_0$, $k_B T =1.0 \epsilon_0$, $\gamma_0=1.0 \, m_0/\tau_0$, $m=m_0$ where $\tau_0$, $\epsilon_0$, $m_0$ and $\sigma_0$ are Lennard-Jones units for time, energy, mass and distance, related as $\epsilon_0 = m_0 \sigma^2_0/\tau_0^2$. The channel is divided in 3 sectors: mouth ($0<x<75\sigma_0$), root $(75\sigma_0<x<700\sigma_0)$ and stomach ($700\sigma_0<x<775\sigma_0$) in every simulation. The right-most section is considered as the stomach and the particles lying there are considered trapped, for counting purposes.

\section*{Acknowledgments}
The authors would like to thank John Chervinsky for assistance in producing the SEM images.  This work was funded by the Rowland Institute at Harvard (RZ and LGW) and the CAPES Science Without Borders Program (CBG, Process Number: 7340/11-7).
C.V. acknowledges fundings  EUR2021-122001, PID2019-105343GB-
I00, IHRC22/00002 and 
PID2022-140407NB-C21 from MINECO.


\begin{thebibliography}{10}
\bibitem{Ellison_PlantBiol06}
A.M. Ellison.
\newblock Nutrient limitation and stoichiometry of carnivorous plants.
\newblock {\em Plant Biol.}, 8:740--747, 2006.

\bibitem{AlbertRJTMDT_JXB10}
V.A. Albert, R.W. Jobson, T.P. Michael, and D.J. Taylor.
\newblock The carnivorous bladderwort ({\it utricularia}, lentibulariaceae): a
  system inflates.
\newblock {\em J. Exp. Bot.}, 61(1):5--9, 2010.

\bibitem{AlbertSWMC_Science92}
V.A. Albert, S.E. Williams, and M.W. Chase.
\newblock Carnivorous plants: Phylogeny and structural evolution.
\newblock {\em Science}, 257:1491--1495, 1992.

\bibitem{Darwin_insect}
C.~Darwin.
\newblock {\em Insectivorous Plants}.
\newblock John Murray, London, 1875.

\bibitem{Fleischmann_book}
A.~Fleischmann.
\newblock {\em Monograph of the Genus {\it Genlisea}}.
\newblock Redfern Natural History Productions Ltd., 2012.

\bibitem{PlachnoMK-KPS_AnnBot07}
B.J. P{\l}achno, M.~Kozieradzka-Kiszkurno, and P.~{\'S}wi\k{a}tek.
\newblock Functional utrastructure [sic] of {\it genlisea} (lentibulariaceae)
  digestive hairs.
\newblock {\em Ann. Bot.-London}, 100:195--203, 2007.

\bibitem{Heslop-Harrison_book}
Y.~Heslop-Harrison.
\newblock Title here.
\newblock In J.T. Dingle and R.T. Dean, editors, {\em Lysosomes in Biology and
  Pathology}, volume~4, pages 525--578. North Holland, Amsterdam, 1975.

\bibitem{PlachnoLAILMPWAJV_PlantBiol06}
B.J. P{\l}achno, L.~Adamec, I.K. Lichscheidl, M.~Peroutka, W.~Adlassnig, and
  J.~Vrba.
\newblock Fluorescence labelling of phosphatase activity in digestive glands of
  carnivorous plants.
\newblock {\em Plant Biol.}, 8(6):813--820, 2006.

\bibitem{PlachnoJFAJ_ActaBotGall05}
B.J. P{\l}achno, J.~Faber, and A.~Jankun.
\newblock Cuticular discontinuities in glandular hairs of genlisea st.-hil. in
  relation to their functions.
\newblock {\em Acta Bot. Gallica}, 152(2):125--130, 2005.

\bibitem{Juniper_book}
B.E. Juniper, R.J. Robins, and D.M. Joel.
\newblock {\em The Carnivorous Plants}.
\newblock Academic Press, London, 1989.

\bibitem{BarthlottSPEFBG_Nature98}
W.~Barthlott, S.~Porembski, Eberhard Fischer, and Bj{\" o}rn Gemmel.
\newblock First protozoa-trapping plant found.
\newblock {\em Nature}, 392(6675):447, 1998.

\bibitem{DarnowskiSF_CPN10}
D.W. Darnowski and S.~Fritz.
\newblock Prey preference in genlisea small crustaceans, not protozoa.
\newblock {\em Carnivorous Plant Newsletter}, 39(4):114--116, 2010.

\bibitem{Petridou2021}
NI~Petridou NI, B~Corominas-Murtra, CP~Heisenberg, and E~Hannezo.
\newblock Rigidity percolation uncovers a structural basis for embryonic tissue
  phase transitions.
\newblock {\em Cell}, 184:1914, 2021.

\bibitem{Mashaghi2014}
A~Mashaghi and C~Dekker.
\newblock Systems and synthetic biology approaches to cell division.
\newblock {\em Syst Synth Biol.}, 8:173, 2014.

\bibitem{Allen2019}
RJ~Allen and B~Waclaw.
\newblock Bacterial growth: a statistical physicist's guide.
\newblock {\em Rep Prog Phys.}, 82:016601, 2019.

\bibitem{Ramaswamy2010}
S.~Ramaswamy.
\newblock The mechanics and statistics of active matter.
\newblock {\em Annual Review of Condensed Matter Physics}, 1:323, 2010.

\bibitem{Ramaswamy2017}
S.~Ramaswamy.
\newblock {\em Journal of Statistical Mechanics: Theory and Experiment},
  70:054002, 2017.

\bibitem{Bechinger2016}
C.~Bechinger, R.~D. Leonardo, H.~Lowen, C.~Reichhardt, G.~Volpe, and G.~Volpe.
\newblock {\em Reviews of Modern Physics}, 88:045006, 2016.

\bibitem{Vicsek2012}
T.~Vicsek and A.~Zafeiris.
\newblock {\em Physics reports}, 517:71, 2012.

\bibitem{Marchetti2013}
M~Cristina Marchetti, Jean-Fran\c{c}ois Joanny, Sriram Ramaswamy, Tanniemola~B
  Liverpool, Jacques Prost, Madan Rao, and R~Aditi Simha.
\newblock {\em Reviews of modern physics}, 85:1143, 2013.

\bibitem{Yan2016}
J.~Yan, M.~Han, J.~Zhang, C.~Xu, E.~Luijten, and S.~Granick.
\newblock {\em Nature materials}, 15:1095, 2016.

\bibitem{vanKesteren2023}
S~van Kesteren, L~Alvarez, S~Arrese-Igor, and L~Isa A~Alegria.
\newblock {\em Proceedings of the National Academy of Sciences},
  120:e2213481120, 2023.

\bibitem{Calero2020}
C.~Calero, J.~Garcia-Torres, A.~Ortiz-Ambriz, F.~Sagues, I.~Pagonabarraga, and
  P.~Tierno.
\newblock {\em Soft Matter}, 16:6673, 2020.

\bibitem{vanderLinden2020}
M.~N. van~der Linden, L.~C. Alexander, D.~G. A.~L. Aarts, and O.~Dauchot.
\newblock {\em Phys.Rev. Lett.}, 123:098001, 2019.

\bibitem{Vutukuri2020}
H.~R. Vutukuri, M.~Lisicki, E.~Lauga, and J.~Vermant.
\newblock {\em Nat. Commun.}, 11:2628, 2020.

\bibitem{Palacci2014}
J.~Palacci, S.~Sacanna, S.-H. Kim, G.-R. Yi, D.~J. Pine, and P.~M. Chaikin.
\newblock {\em Philos.Trans. R. Soc. A}, 372:20130372, 2014.

\bibitem{Manson2020}
Manson MD.
\newblock Howard berg's random walk through biology.
\newblock {\em J Bacteriol}, 202:e00494, 2020.

\bibitem{Cates2013}
M.~E. Cates and J.~Tailleur.
\newblock {\em Europhys. Lett}, 101:20010, 2013.

\bibitem{Qiu2014}
Tian Qiu, Tung-Chun Lee, Andrew~G. Mark, Konstantin~I. Morozov, Raphael
  Munster, Otto Mierka, Stefan Turek, Alexander~M. Leshansky, and Peer Fischer.
\newblock {\em Nature Communications}, 5:5119, 2014.

\bibitem{BergDB_Nature72}
H.~C. Berg and D.~A. Brown.
\newblock Chemotaxis in {\it escherichia coli} analysed by three-dimensional
  tracking.
\newblock {\em Nature}, 239:500, 1972.

\bibitem{ThorntonJBSDBBLW_NatComms20}
K.L. Thornton, J.K. Butler, S.J. Davis, B.K. Baxter, and L.G. Wilson.
\newblock Haloarchaea swim slowly for optimal chemotactic efficiency in low
  nutrient environments.
\newblock {\em Nat. Commun.}, 11:4453, 2020.

\bibitem{XieTASCX-LW_PNAS10}
L.~Xie, T.~Altindal, S.~Chattopadhyay, and X.-L. Wu.
\newblock Bacterial flagellum as a propeller and as a rudder for efficient
  chemotaxis.
\newblock {\em Proc. Natl. Acad. Sci. USA}, 108:2246--2251, 2011.

\bibitem{SonJGRS_NatPhys13}
K.~Son, J.S. Guasto, and R.~Stocker.
\newblock Bacteria can exploit a flagellar buckling instability to change
  direction.
\newblock {\em Nat. Phys.}, 9:494--498, 2013.

\bibitem{ArmitageRS_Microbiol97}
J.P. Armitage and R.~Schmitt.
\newblock Bacterial chemotaxis: {\it {R}hodobacter sphaeroides} and {\it
  {s}inorhizobium meliloti} - variations on a theme?
\newblock {\em Microbiol.}, 143:3671--3682, 1997.

\bibitem{galajda2007wall}
Peter Galajda, Juan Keymer, Paul Chaikin, and Robert Austin.
\newblock A wall of funnels concentrates swimming bacteria.
\newblock {\em Journal of bacteriology}, 189(23):8704--8707, 2007.

\bibitem{ReichardtCR_PRL06}
C~Reichhardt and C~J~Olsen Reichhardt.
\newblock Crossover from intermittent to continuum dynamics for locally driven
  colloids.
\newblock {\em Phys. Rev. Lett.}, 96:028301, 2006.

\bibitem{Martinez2018}
Raul Martinez, Francisco Alarcon, Diego~Rogel Rodriguez, Juan~Luis Aragones,
  and Chantal Valeriani.
\newblock {\em The European physical journal. E, Soft matter}, 41:91, 2018.

\bibitem{vanHoutenEMTK_JProtozool82}
J.~van Houten, E.~Martel, and T.~Kasch.
\newblock Fluorescence labelling of phosphatase activity in digestive glands of
  carnivorous plants.
\newblock {\em J. Protozool.}, 29(2):226--230, 1982.

\bibitem{FOISSNER2014}
W.~Foissner.
\newblock Protozoa.
\newblock In {\em Reference Module in Earth Systems and Environmental
  Sciences}. Elsevier, 2014.

\bibitem{RaynaudNN_PLoSOne14}
X.~Raynaud and N.~Nunan.
\newblock Spatial ecology of bacteria at the microscale in soil.
\newblock {\em PLoS One}, 9:e87217, 2014.

\bibitem{LundbergSLSPSYJGSMJTAEVKTdRRETERLPHSTJD_Nature12}
D.S. Lundberg, S.L. Lebeis, S.H. Paredes, S.~Yourstone, J.~Gahring,
  S.~Malfatti, J.~Tremblay, A.~Engelbrekston, V.~Kunin, T.~Glavina del Rio,
  R.C. Edgar, T.~Eickhorst, R.E. Ley, P.~Hugenholtz, S.~Green Tringe, and J.L.
  Dangl.
\newblock Defining the core {\it arabidopsis thaliana} root microbiome.
\newblock {\em Nature}, 488:86--90, 2012.

\bibitem{TurnerLPMNHB_BiophysJ16}
L.~Turner, L.~Ping, M.~Neubauer, and H.C. Berg.
\newblock Visualizing flagella while tracking bacteria.
\newblock {\em Biophys. J.}, 111:630--639, 2016.

\bibitem{ReichhardtCR_PRE04}
C.~Reichhardt and C.~J. Olson~Reichhardt.
\newblock Directional locking effects and dynamics for particles driven through
  a colloidal lattice.
\newblock {\em Phys. Rev. E}, 69(4):041405, Apr 2004.

\bibitem{Martinez2020}
Raul Martinez, Francisco Alarcon, Juan~Luis Aragones, and Chantal Valeriani.
\newblock {\em Soft Matter}, 16:4739, 2020.

\bibitem{PlachnoKAJFJK_ActaBotGall05}
B.J. P{\l}achno, K.~Adamus, J.~Faber, and J.~Koz{\l}owski.
\newblock Feeding behaviour of carnivorous genlisea plants in the laboratory.
\newblock {\em Acta Bot. Gallica}, 152(2):159--164, 2005.

\bibitem{Heslop-Harrison_Chapter}
Y.~Heslop-Harrison.
\newblock Unknown.
\newblock In J.T. Dingle and R.T. Dean, editors, {\em Lysosomes in Biology and
  Pathology}, volume~4, pages 525--578. North Holland, Amsterdam, 1975.

\bibitem{GlisteningCarnivores}
Stewart McPherson.
\newblock {\em Glistening Carnivores --- The Sticky-Leaved Insect-Eating
  Plants}.
\newblock Redfern Natural History Productions, 2008.

\bibitem{Perrin_book}
J.~Perrin.
\newblock {\em Atoms}.
\newblock D. Van Nostrand Company, New York, 1916.

\bibitem{WilsonRZ_OptExp12}
Laurence Wilson and Rongjing Zhang.
\newblock 3d localization of weak scatterers in digital holographic microscopy
  using rayleigh-sommerfeld back-propagation.
\newblock {\em Opt. Express}, 20(15):16735--16744, 2012.

\bibitem{WilsonVMJS-LJTGBPPWP_PRL11}
L.~G. Wilson, V.~A. Martinez, J.~Schwarz-Linek, J.~Tailleur, G.~Bryant, P.~N.
  Pusey, and W.~C.~K. Poon.
\newblock Differential dynamic microscopy of bacterial motility.
\newblock {\em Phys. Rev. Lett.}, 106(1):018101, Jan 2011.

\end{thebibliography}
\end{document}